\documentclass[fleqn,usenatbib]{mnras}
\usepackage{newtxtext,newtxmath}

\usepackage[T1]{fontenc}

\usepackage{mathabx}

\DeclareRobustCommand{\VAN}[3]{#2}
\let\VANthebibliography\thebibliography
\def\thebibliography{\DeclareRobustCommand{\VAN}[3]{##3}\VANthebibliography}

\usepackage{graphicx}	
\usepackage{amsmath}	
\newcommand{\angstrom}{\textup{\AA}}

\title[The ultraviolet habitable zone of exoplanets]{The ultraviolet habitable zone of exoplanets}

\author[R.~Spinelli et al.]{
R.~Spinelli$^{1}$\thanks{E-mail: rspinelli@uninsubria.it}
F.~Borsa$^{2}$
G.~Ghirlanda$^{2,3}$
G.~Ghisellini$^{2}$
F.~Haardt$^{1,2,3}$
\\
$^{1}$Dipartimento di Scienza e Alta Tecnologia, Universit\`a dell'Insubria, Via Valleggio 11, 22100 Como, Italy\\
$^{2}$ INAF -- Osservatorio Astronomico di Brera, Via E. Bianchi 46, 23807 Merate (LC), Italy\\
$^{3}$INFN -- Sezione Milano--Bicocca, Piazza della Scienza~3, 20126 Milano, Italy
}

\date{Accepted 2023 March 23. Received 2023 February 24; in original form 2022 October 13}

\pubyear{2023}

\begin{document}
\label{firstpage}
\pagerange{\pageref{firstpage}--\pageref{lastpage}}
\maketitle

\begin{abstract}
The dozens of rocky exoplanets discovered in the Circumstellar Habitable Zone (CHZ) currently represent the most suitable places to host life as we know it outside the Solar System. However, the presumed presence of liquid water on the CHZ planets does not guarantee suitable environments for the emergence of life. According to experimental studies, the building blocks of life are most likely produced photochemically in presence of a minimum ultraviolet (UV) flux. On the other hand, high UV flux can be life-threatening, leading to atmospheric erosion and damaging biomolecules essential to life. These arguments raise questions about the actual habitability of CHZ planets around stars other than Solar-type ones, with different UV to bolometric luminosity ratios. By combining the "principle of mediocricy" and recent experimental studies, we define UV boundary conditions (UV-habitable Zone, UHZ) within which life can possibly emerge and evolve. We investigate whether exoplanets discovered in CHZs do indeed experience such conditions. By analysing Swift-UV/Optical Telescope data, we measure the near ultraviolet (NUV) luminosities of 17 stars harbouring 23 planets in their CHZ.  We derive an empirical relation between NUV luminosity and stellar effective temperature. We find that eighteen of the CHZ exoplanets actually orbit outside the UHZ, i.e., the NUV luminosity of their M–dwarf hosts is decisively too low to trigger abiogenesis – through cyanosulfidic chemistry – on them. Only stars with effective temperature $\gtrsim$3900 K illuminate their CHZ planets with enough NUV radiation to trigger abiogenesis. Alternatively, colder stars would require a high-energy flaring activity. 

\end{abstract}



\section{Introduction}
In the last 20 years exoplanet research has made great strides. More than 5000 exoplanets have been discovered in our Galaxy and more than 8000 candidates are awaiting confirmation from further observations\footnote{\url{https://exoplanets.nasa.gov/}}. After an hectic discovery epoch, the research activities are now starting to focus on the characterization of exoplanets with the aim of understanding the planetary population around us and identifying exoplanets with physical properties suitable for life.

Exoplanets around M--dwarf stars may represent the most promising targets to discover habitable worlds. 
Besides being the most abundant stellar population in our Galaxy ($\sim$75\%, \citealt{Bochansky2010}), M-dwarfs are extremely long-lived and thus offer long timescales for possible biological evolution within their planetary systems \citep{Shields2016}. 
Moreover, the physical properties of M-dwarfs, as their small radii, low masses and low luminosities favour their habitable zone exoplanets \citep[][]{Kasting1993, Kopparapu2013} discovery through methods based on transits and radial velocities \citep{Nutzman2008, Quirrenbach2014}. 
Eventually, M--dwarf habitable zone exoplanets will allow us to perform a characterization of their atmospheres  \citep[e.g.,][]{Snellen2013, Rodler2014, Batalha2015, Barstow2016, Lusting2019} through spectroscopy by the James Webb Space Telescope \citep{Gardner2006} and Extremely Large Telescopes \citep{Gilmozzi2007,dezeeuw2014}.

These studies will possibly include the identification of key biomarkers, such as O$_2$, O$_3$, CH$_4$, and N$_2$O, whose maintenance on Earth is supported by the presence of life. 
How to interpret an eventual detection of these biomarkers in the exoplanetary atmospheres is the subject of several works \citep[e.g.,][]{Owen1980, Sagan1993, Desmarais2002, Catling2018, Schwieterman2018}. 

On the theoretical side, it is worth investigating whether life can emerge and endure on such worlds. 
Several factors contribute to the emergence and development of life (at least as we know it) on planets. 
Given its importance for life on Earth, liquid water is considered a necessary factor. 
Thus a rocky planet is considered potentially habitable if it resides in the so-called circumstellar habitable zone \citep[CHZ, e.g.,][]{Kasting1993, Kopparapu2013}, 
namely the annular region around a star in which there are suitable temperatures 
for the presence of liquid water on the planet's surface. 
This region is primarily determined by the star-planet distance and the stellar 
properties (such as luminosity and effective temperature $T_{\rm eff}$), but it can also depend 
on other factors such as atmospheric composition \citep{Zsom2013}, planetary mass \citep{Kopparapu2014}, volcanic activity \citep{Ramirez2017} and tidal locking \citep{Driscoll2015, Barnes2017}. 

In addition to the presence of liquid water,  other factors may contribute to making a planet habitable \citep{Meadows2018}.
For example, intrinsic properties of the planet (e.g., ongoing plate tectonics and volcanism) play a crucial role in  temperature regulation and may generate  magnetic fields that limit atmospheric stripping due to the stellar wind \citep[e,g.,][]{Lenardic2014, Tosi2017, Godolt2019, McIntyre2022}. 

The  radiation environment, determined by the host star and/or other galactic powerful 
sources, may influence habitability. 
For example M-dwarfs are known to be strong and variable X-ray/UV emitters as a consequence 
of magnetic activity in their outer atmospheres \citep{France2012, Stelzer2013}. 
In particular, the high energy UV/X--ray emission of the host star may have a crucial role 
for the habitability of its closest ($\sim$0.1 AU) planets. 
The effect of the high energy radiation environment on habitability could be twofold. 
UV/X-ray radiation may cause atmospheric erosion \citep{SanzForcada2010}, biomolecules destruction \citep{Sag} and damage to various species of proteins and lipids \citep{Buccino2007}. 
On the larger galactic length scales (pc to kpc), planetary habitability can also be threatened by powerful astrophysical transient of high energy radiation, like Gamma Ray Bursts (GRBs) and Supernovae (SNe), which can trigger mass extinction events \citep{Ruderman1974,Thorsett1995,Gehrels2004,Melott2011, Piran2014, Spinelli2021}. 
On the other hand,  experimental studies \citep[e.g.,][]{Toupance1977, Powner2009, Ritson2012, Patel2015, Xu2018, Rimmer2018} demonstrate that UV light is a crucial ingredient for prebiotic photochemistry, namely for the synthesis of ribonucleic acid (RNA), i.e. the building blocks for the emergence of life. 

In this work we study and combine the positive (prebiotic) and negative (atmospheric erosion and biomolecules destruction) roles of stellar UV radiation on potentially habitable exoplanets. We define the Ultraviolet Habitable Zone (UHZ) in \S2 and compare with a sample of exoplanets, selected in \S3, for which we evaluate (\S3) the UV luminosity of the host star by exploiting the existing public observations of the Ultra-Violet Optical Telescope (UVOT - \citealt{Roming2005}) on board the Neil Gehrels {\it Swift} Observatory \citep{Gehrels2004}. Results are presented in \S4 where we compare the UV habitable zone with the UV luminosity of selected exoplanets and with their habitable zone as defined by the classical H$_2$O criterium \citep{Kopparapu2013}. Results are discussed in \S5. 

\section{UV habitable zone}
\label{sec:habitablezone}

Assuming that the conditions for the existence of life in the Universe are similar to those on Earth is the "hard core" \citep{Lakatos1974} of all programs in search of life in the Universe. Therefore, among all the exoplanets discovered, rocky planets with liquid water on their surface are considered the best candidates for finding life outside the Solar System. For this reason, the search for life forms outside the Solar System focuses on this particular class of exoplanets.

While the presence of liquid water is considered a necessary condition for the maintenance of life, how life actually stems from prebiotic conditions is one of the greatest open questions for the understanding of our very existence. Several studies \citep[see, e.g.,][]{Patel2015, 2017ApJ...843..110R, Xu2018} suggest scenarios in which the intensity of UV radiation reaching a planet's surface may play a key role in the emergence of life. 

This holds only for scenarios that require UV light. Moreover, the specific thresholds depends on the scenario and chemical reactions considered \citep[see][]{Rimmer2021,2017ApJ...843..110R}.
The contrasting effects of UV radiation on life, i.e., the action that favours prebiotic activity versus the atmospheric erosion and DNA destruction,  define two boundaries that delimit a region around a star where exoplanets are subject to favourable conditions in terms of UV irradiation. We therefore define this region as the {\it UV--habitable zone}, or UHZ.

Inspired by the so-called “Principle of Mediocrity” \citep{Vonhoerner1961}, i.e., the idea that the conditions that allowed the origin and evolution of life on Earth are average in comparison with other inhabited worlds in the Universe,  \citet{Buccino2007} defined the UHZ as follows. 
The boundaries of the UHZ are calculated by considering the intensity of UV radiation reaching the Archean Earth (AE) 3.8 Gyrs ago, the era of the first traces of life on Earth \citep{Dodd2017}, which we indicate as $S(AE)_{\rm UV}$. Notice that this symbol refers to the flux incident above the AE atmosphere.  
With respect to the planet--star separation, \citet{Buccino2007} defines the inner UHZ boundary as the distance corresponding to the maximum tolerable dose of UV radiation for biological systems, i.e., $S_{\rm UV}<2\,S(AE)_{\rm UV}$, and the outer UHZ boundary as the distance corresponding to the  minimum flux needed for the chemical synthesis of complex molecules (such as amino acids, lipids and nucleosides), i.e., $S_{\rm UV}>0.5\, S(AE)_{\rm UV}$. These fluxes are then re-scaled accounting for possible absorption by the planet atmosphere to compute the actual flux incident on the planet's surface. 


Given their ability to combines with other molecules in aqueous solution and create to important biomolecules for the origin of RNA, hydrogen cyanide (HCN) is considered a fundamental molecules for the origin of life in the context of the RNA world hypothesis  \citep{Rich1962, Gilbert1986}. Among these molecules there are amino acid glycine \citep{Miller1957}, nucleobase adenine \citep{Ferris1966, Sanchez1966, Sanchez1970}, lipids and nucleosides \citep{Ritson2012, Patel2015,Xu2018}. 

Recently, \citet{Rimmer2018} showed that, by illuminating a mixture of HCN and SO$^{2-}_{3}$ with Near-UV (NUV, 200-280 nm) light, RNA pyrimidine nucleotide precursors are produced. 
On the contrary, the absence of NUV light in the same mixture leads to the formation of inert adducts with no prebiotic potential. \citet{Rimmer2018} defined the minimum NUV flux required for abiogenesis as the flux which triggers a 50\% yield of the photochemical product at temperature of 0 \textdegree C, in the SO$^{2-}_{3}$ reaction. 
This corresponds to integrate a specific flux of 6.8 $\times$ 10$^{9}$ photons cm$^{-2}$ s$^{-1}$ \AA$^{-1}$ over 
the 200 - 280 nm band. 

In the present work we consider the, experimentally motivated, threshold flux of \citet{Rimmer2018}, i.e., $S_{\rm UV}>45$ erg cm$^{2}$ s$^{-1}$, defining the outer boundary of the UHZ. 
For the maximum tolerable flux, setting the inner UHZ boundary, we consider the criterion of \citet{Buccino2007}, i.e., $S_{\rm UV}<2\times S(AE)_{\rm UV}=2\times5.2\times 10^3$ erg cm$^{-2}$ s$^{-1}$, where for $S(AE)_{\rm UV}$ we adopt the solar spectra of \citet{Thuillier2004} and assume that the Sun, during the Archean Eon, was radiating, in the entire 200-280 nm wavelength range, at 75\% of its present-day power \citep{Cockell1998, Cockell2000}. Finally, in order to account for atmospheric absorption, we consider that the flux that actually reaches a planet's surface can be 10\%, 50\% or 100\% of that at the top of the atmosphere (respectively a transmission $f$ of 0.1, 0.5, 1.0). 
Here  we do not aim at any frequency-dependent description of atmospheric UV transmission, rather we consider an average value over the relevant frequency range.


\section{Sample selection and data analysis}
\label{sec:sample}

The aim of the present study is to set the UHZ in the context of exoplanets considered potentially habitable based solely on the classical H$_2$O criterion, i.e., planets orbiting their host stars in the CHZ. Our targets should satisfy three conditions: (i) be rocky planets, (ii) reside in the CHZ of their host stars for which (iii) NUV observations are available. 

\subsection{Sample selection}

We consider the catalog of potentially habitable planets provided by the Planetary Habitability Laboratory (PHL, University of Puerto Rico, Arecibo).\footnote{\url{https://phl.upr.edu/the-habitable-exoplanets-catalog}}\footnote{The PHL catalog is complete up to 59. Therefore, we updated it to June 2022 by applying the PHL selection criteria to the NASA Exoplanet Archive \url{https://exoplanetarchive.ipac.caltech.edu/}. We find 14 more exoplanets with no Swift UVOT observations in the considered filters.} The PHL selects planets with  $0.5 \le R_P\le 1.6 $ or with  
$0.1 \le M_P\sin{i} \le 3$ and planets with $1.6 \le R_P\le 2.5$ or with  
$3 \le M_P \sin{i}\le 10$ (planetary radii $R_P$ and masses $M_P$ are in Terrestrial units $R_{\Earth}$ and $M_{\Earth}$). 

The first set of conditions selects Earth--like planets (Conservative Sample, CS) of probable rocky composition, while the second one widens the sample to ocean worlds and mini-Neptunes (Optimistic Sample, OS).  
The lower limit of 0.5 R$_{\Earth}$ (or 0.1 M$_{\Earth}$) excludes planets not massive enough to retain their atmospheres \citep{Zahnle2017}.

\begin{figure} 
    \centering
    \includegraphics[width=0.50\textwidth]{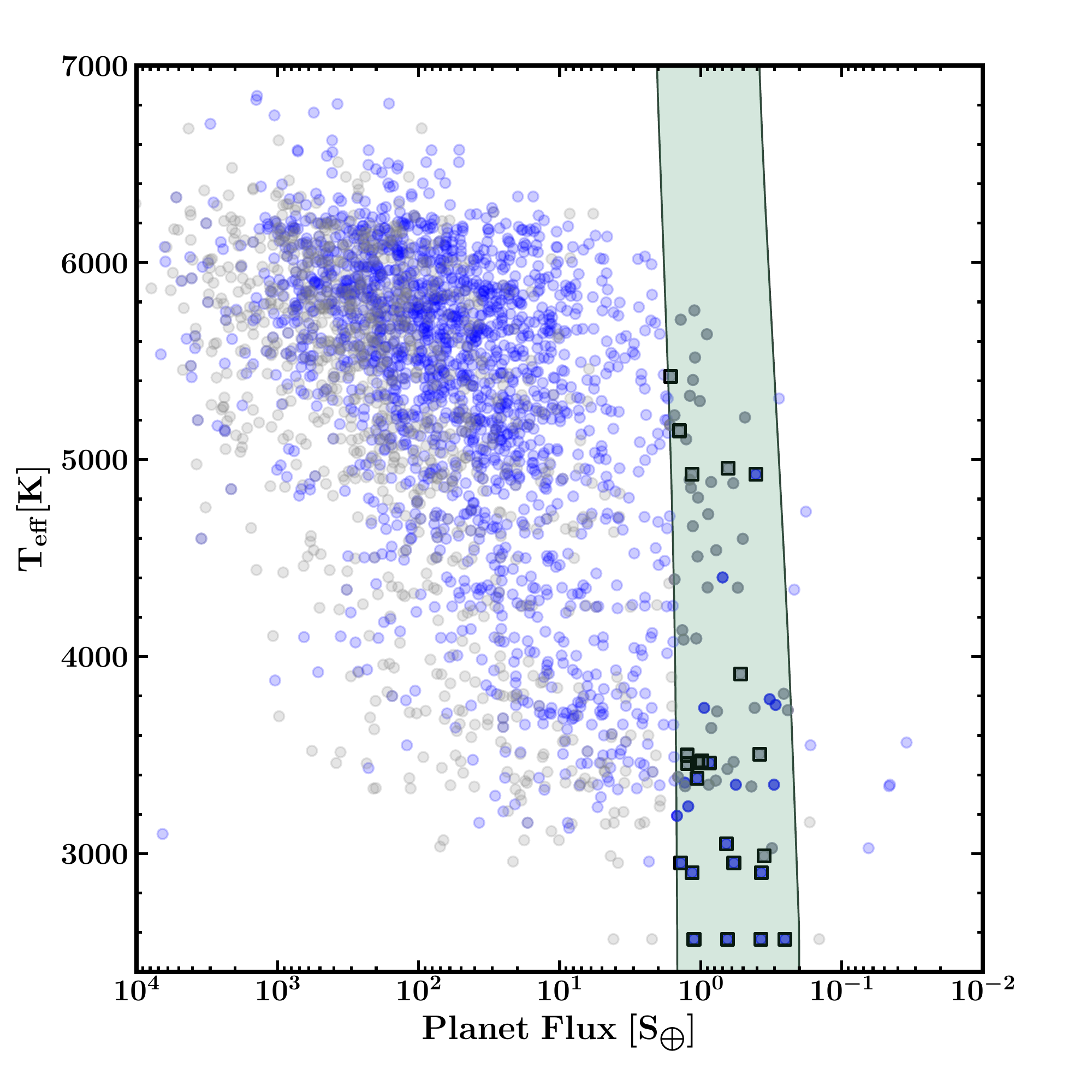}
    \caption{$T_{\rm eff}$ versus planet received flux (in unit of the flux received today by the earth $S_\Earth=1.36 \times 10^6$ erg cm$^{-2}$ s$^{-1}$) above the atmosphere. Grey symbols: discovered planets with  
$0.5 \le R_P\le 1.6 $ or with  
$0.1 \le M_P\sin(i) \le 3$ and planets with $1.6 \le R_P\le 2.5$ or with  
$3\le M_P \sin(i)\le 10$ (radii and masses in Terrestrial units). Green zone: the optimistic habitable zone as defined by \citet{Kopparapu2013}. Square symbols: planets with archival {\it Swift}/UVOT observations in UM2/UW2/UW1 filters.}
    \label{fig:sample}
\end{figure}

From the host star luminosity and semi-major axis reported in the PHL we compute the bolometric insulation received by each planet. Figure \ref{fig:sample} shows all planets of the PHL sample in the stellar $T_{\rm eff}$-insulation plane (CS and OS planets are shown by the grey and blue symbols, respectively). 

Next, we need to select those planets which are in the CHZ 
\citep[see][for different definitions for the CHZ boundaries]{Kopparapu2013,Kopparapu2014}.  
In this work we adopt the optimistic limits of the CHZ, empirically determined by \citet{Kasting1993} and \citet{Kopparapu2013}. These limits are based on the inferred presence of liquid water on Mars' surface before 3.8 Gyr ago \citep{Pollack1987,Bibring2006}, and on the absence of liquid water on Venus' surface for, at least, the past Gyr \citep{Solomon1991}. It is assumed (optimistically) that Venus and Mars actually had liquid water on 
their surfaces, respectively 1 and 3.8 Gyr ago and, therefore, that they were habitable at those epochs.

In particular, \citet{Kopparapu2013} estimate the inner edge of the solar CHZ by computing the solar irradiation experienced by Venus 1 Gyr ago when the Sun luminosity was 92\% of its present--day value. This flux, computed at Venus--Sun distance, corresponds to 1.76 times the flux received today by the Earth ($S_{\Earth}=1.36 \times 10^6$ erg cm$^{-2}$ s$^{-1}$). The outer edge of the solar CHZ is set by considering the flux received by Mars 3.8 Gyr ago when the solar luminosity was 75\% of its present--day value, i.e., $\simeq 0.32$ $S_{\Earth}$. These limiting flux values are defined for solar-type stars. For stellar types  other than solar, \citet{Kopparapu2013} provide corrective factors as function of the stellar $T_{\rm eff}$.  

\begin{table}
\centering
	\begin{tabular}{ccccccc
    	}
    	\hline
    	\hline
    	                                                                 %
        Star & Spectral &  $T_{\rm eff}$ & $a$ & $S$ & $d$ &  ESI  \\ 
        & Type & [K] & AU & S$_{\Earth}$ & [pc] & \\


        \cline{1-7}
        Trappist-1 d & M8.0 & 2550 & 0.022 & 1.12& 12.43 & 0.91\\
        Trappist-1 e & M8.0 &  2550  & 0.029 & 0.65 & 12.43 & 0.85 \\
        Trappist-1 f & M8.0 & 2550  & 0.038& 0.37 & 12.43 & 0.68\\
        Trappist-1 g &  M8.0 &2550  & 0.047& 0.25 & 12.43 & 0.58 \\
        Teegarden b   &  M7.0 & 2790 & 0.025 & 1.15 &  3.83& 0.95\\
        Teegarden c   &  M7.0 & 2790 & 0.044  & 0.37 &  3.83 & 0.68\\
        GJ 1061 c  &  M5.5 & 2905& 0.035& 1.45 & 3.67 & 0.86 \\
        GJ 1061 d  &  M5.5 & 2905& 0.054& 0.69 & 3.67 &  0.86 \\
        Proxima Cen b & 	M5.5 & 3050   & 0.049& 0.65& 1.30 & 0.87\\
        LHS 1140 b  &  M4.5 & 3166& 0.096&  0.50 & 14.99  & 0.62 \\
        GJ 273 b & M3.5 & 3382  & 0.091 & 1.06 & 3.79 & 0.85 \\
        GJ 163 c &  M3.5 & 3399 & 0.125& 1.25& 15.13 & 0.69\\
        K2-18 b & M2.5 &  3464 & 0.143 & 1.00 & 38.03 & 0.70  \\
        GJ 357 d & 	M2.5 & 3490& 0.204 & 0.38 & 9.44 & 0.58\\
        TOI-700 d &  M2.5 & 3494 &  0.163 & 0.85 & 31.13 & 0.93 \\
        GJ 832 c &M2.0 & 3601& 0.163 & 0.99& 4.96 & 0.74 \\ 
        GJ 433 d &  M2.0& 3605& 0.178 & 1.03& 9.06 & 0.74 \\
        GJ 229A c &  M0.0 & 3912 & 0.384 &0.53 &5.76 & 0.69\\
        Kepler-62 f &  K2.0 & 4842&  0.427 & 1.35 & 300.87 & 0.68 \\
        Kepler-62 e & K2.0 & 4842& 0.718& 0.48 & 300.87 & 0.83\\
        HD 40307 g &  K2.0 & 4867& 0.600 & 0.67& 12.94& 0.66 \\
        Kepler-1606 b &   G7.0 & 5400 & 0.642 & 1.29 & 831.29 & 0.70\\
        Kepler-1701 b & K1.0 & 5146 & 0.561 & 1.31 & 584.14 & 0.71\\

    	\hline
    	\hline
	\end{tabular}
	\caption{CHZ planets considered in this work, with the spectral type and $T_{\rm eff}$ of the host star, the semi-major axis of the CHZ planet, the flux received by the planet (in the unit of the flux received today by the earth), the distance from the earth and the Earth Similarity Index from the PHL.}
\label{tab:sample}
\end{table}

Figure \ref{fig:sample} shows the CHZ (green shaded region) computed according to the definition of \citet{Kopparapu2013}. In general, the CHZ is weakly dependent on the stellar $T_{\rm eff}$ and it is bracketed by flux values, received by the planet's atmosphere, 0.2--1.4$S_{\Earth}$ for low temperature stars ($T_{\rm eff}<3000$ K) and 0.4--2$S_{\Earth}$ for large $T_{\rm eff}$ values. 

In order to estimate the host star NUV luminosity, we searched into the {\it Swift}/UVOT archive and found observations for 23 of the 73 selected CHZ targets (grey and blue squares in Fig.\ref{fig:sample}). This is the final selected sample of planets in the CHZ with UV observations. A list of the corresponding planetary parameters is provided in Table \ref{tab:sample}.

\subsection{Swift/UVOT data analysis}
\label{sec:analysis}
{\it Swift} is a space observatory mainly dedicated to the study of Gamma-Ray Burst (GRB). {\it Swift} obtains X--ray and UV  data simultaneously through the co-aligned X--Ray Telescope (XRT, 0.2-10 keV - \citealt{Burrows2005}) and Ultra Violet Optical Telescope (UVOT, 1700-6500 $\AA$ - \citealt{Roming2005}).  
For each source in our sample we collected from the {\it Swift} archive all the archival UVOT observations performed with any of the following filters: UM2 ($\lambda_{\rm eff}$=2231 $\angstrom$, FWHM=498 $\angstrom$) and
UW2 ($\lambda_{\rm eff}$=2030 $\angstrom$, FWHM=657 $\angstrom$). 
UVOT images (taken with the UVW2, UVM2 filters) were analyzed with the public \texttt{HEASOFT (v6.30)} software package\footnote{\url{https://heasarc.gsfc.nasa.gov/docs/software/heasoft/}}. The most recent version of the calibration database was used\footnote{\url{https://heasarc.gsfc.nasa.gov/docs/heasarc/caldb/}}. 
Photometry was performed within a circular source-extraction region of 5 arcsec in radius centered on the source, with  
the background extracted from a circular region within a radius of about 20 arcsec, close to the target but free from contamination. For each source we stacked all the observations using the \texttt{uvotimsum} tool. We adopted a $3 \sigma$ detection threshold. 
Source count rates were then converted into  flux densities using the conversion factors given in Table 1 of \citet{Brown2016}, valid for stellar spectra as given by \citet{Pickles1998}. 

When available, we analysed and preferably considered data from UM2 filter as it best covers the NUV range. In case no UM2 data were available, we considered UW2 data. 
In this case, special care was needed as the transmission curve of UW2 is characterised by a red-tail extending in the optical \citep{Breeveld2011}. As a consequence, the integrated flux could be contaminated by the (much brighter) black body optical emission from the star's photosphere. In order to correct for such contamination, we adopted the red-tail magnitude corrections as given in Table 12 of \citet{Brown2010}, which allowed us to disentangle the genuine UV emission from the contaminant stellar black body emission. In table \ref{tab:log1} we report the total exposure time, filter and the average count rate used in this work for each star. For uw2 filter the count rates are already corrected for the red tail contamination.

\begin{table} 
\renewcommand{\arraystretch}{1}
\centering
	\begin{tabular}{cccc}
 \hline
\hline
  \multicolumn{1}{c}{Source} &
  \multicolumn{1}{c}{Filter} &
 \multicolumn{1}{c}{Exp. time [ks]}&
 \multicolumn{1}{c}{Counts s$^{-1}$}\\
 \hline
Trappist-1 & uw2 & 274 & (5.32 $\pm$ 1.33) $\times$ 10$^{-6}$\\
Teegarden's star & uw2 & 34 & (2.75 $\pm$ 0.15) $\times$ 10$^{-4}$\\ 
GJ 1061 & um2& 36 & 0.016 $\pm$ 0.002 \\ 
Proxima Cen & um2& 77 & 1.127 $\pm$ 0.038\\ 
LHS 1140 & uw2& 83 & 0.0314$\pm$ 0.0008\\ 
GJ 273 & um2 & 30 & 0.223 $\pm$ 0.002\\  
GJ 163 & um2 & 6 & 0.065 $\pm$ 0.006\\ 
K2-18 & uw2& 12& 0.0175 $\pm$ 0.0005  \\ 
GJ 357 & uw2& 16 & 0.1416 $\pm$ 0.0003 \\
TOI-700 & um2 & 17 & 0.0096 $\pm$ 0.0026\\
TOI-700 & uw2 & 22 & 0.015 $\pm$ 0.0002 \\ 
GJ 832 & um2& 39 & 0.735 $\pm$ 0.016\\
GJ 433 & um2 & 6 & 0.297 $\pm$ 0.011\\
GJ 229 A & um2 & 0.5 & 2.304 $\pm$ 0.090 \\
GJ 229 A & um2 & 0.6 &3.569 $\pm$ 0.076 \\
Kepler-62 & uw2& 5 &  0.0266 $\pm$ 0.0018\\
HD 40307 & um2& 1.5 & 16.645 $\pm$ 0.368 \\
Kepler-1606 & um2& 0.7& 0.046 $\pm$ 0.012\\
Kepler-1701 & um2& 1.7 &0.036 $\pm$ 0.009\\
\hline
\hline
\end{tabular}\caption{Stars considered in this work with the filter, the total exposure time and the average count rate of the analyzed observations. For the uw2 filter the count rates are already corrected for the red tail contamination as described in \S \ref{sec:analysis}.}
\label{tab:log1}
\end{table}


%

%
\section{Results}

\begin{figure*}
    \centering
    \includegraphics[width=0.90\textwidth]{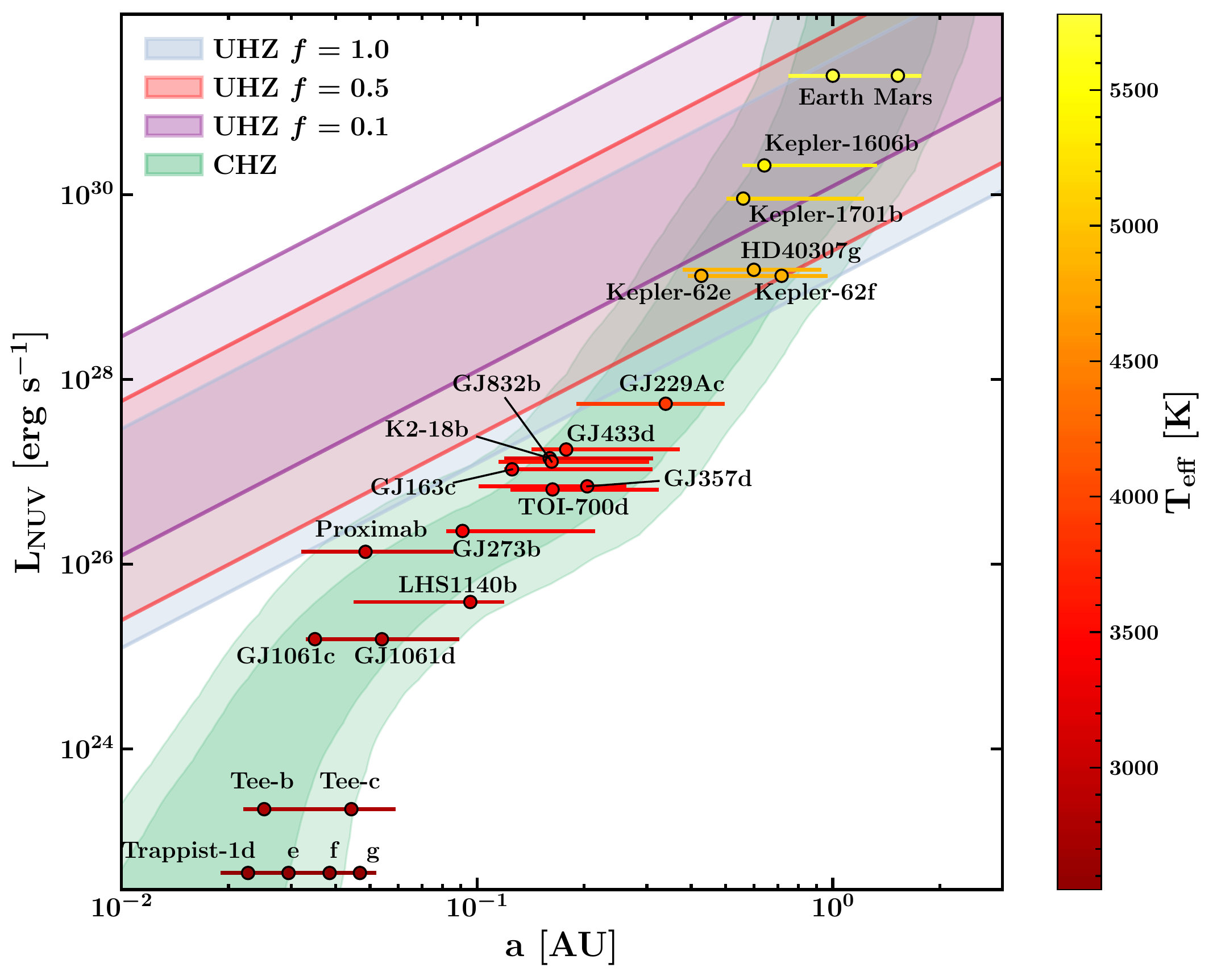}
    \caption{NUV (Near Ultra-Violet) luminosity versus star--planet separation. The UHZ (Ultraviolet Habitable zone, see \S2) is defined by the inner (outer) boundary corresponding to the maximum (minimum) UV flux tolerable required) for life resilience (abiogenenesis, considering a specific prebiotic pathways). Three UHZ are shown corresponding to 100, 50 and 10\% (as labelled) atmospheric transmission ($f$), to the planet's surface, of the UV flux at the top of the atmosphere. The planets of our sample (\S3) selected for being in the CHZ (Circumstellar Habitable Zone) and having UV observations (thus providing $L_{\rm NUV}$) are shown by the circle symbols. The color coding corresponds to the $T_{\rm eff}$ of their host stars (vertical color bar). The horizontal bar of each exoplanet shows the CHZ as derived by \citet{Kopparapu2013}. The green shaded regions show the CHZ as derived through the $L_{\rm NUV}-T_{\rm eff}$ correlation (\S5).  }
    \label{fig:NUV}
\end{figure*}

The NUV luminosity $L_{\rm NUV}$ for all the stars in our sample were estimated through the stellar distances of GAIA DR2 \citep{Gaia2018}. To this aim we assumed a flat spectrum (in erg cm$^{-2}$ s$^{-1}$ \AA$^{-1}$) in the NUV band, as supported by
visual inspection of archival NUV data obtained by HST/STIS and spectra delivered by the treasury MUSCLES survey \citep{2016ApJ...820...89F}. 

Fig.\ref{fig:NUV} shows the NUV luminosity $L_{\rm NUV}$ versus the star--planet separation $a$. In this plane  the UHZ is represented as  a stripe, bounded by the inner and outer values defined in \S~\ref{sec:habitablezone}.
We considered three different NUV atmospheric transmission values: 10\% (violet), 50\% (red) and 100\% (grey). Our targets are shown (circles) in the $L_{\rm NUV}-a$ according to the NUV luminosity we derived from {\it Swift}-UVOT data analysis. The horizontal line on each planetary system shows the extension of the relative CHZ. The latter is derived as in \citet{Kopparapu2013} from the star's bolometric luminosity and $T_{\rm eff}$. Fig.\ref{fig:NUV} allows us to compare 
the CHZ of specific stellar systems to the UHZ. 

\begin{figure} 
    \centering
    \includegraphics[width=0.52\textwidth]{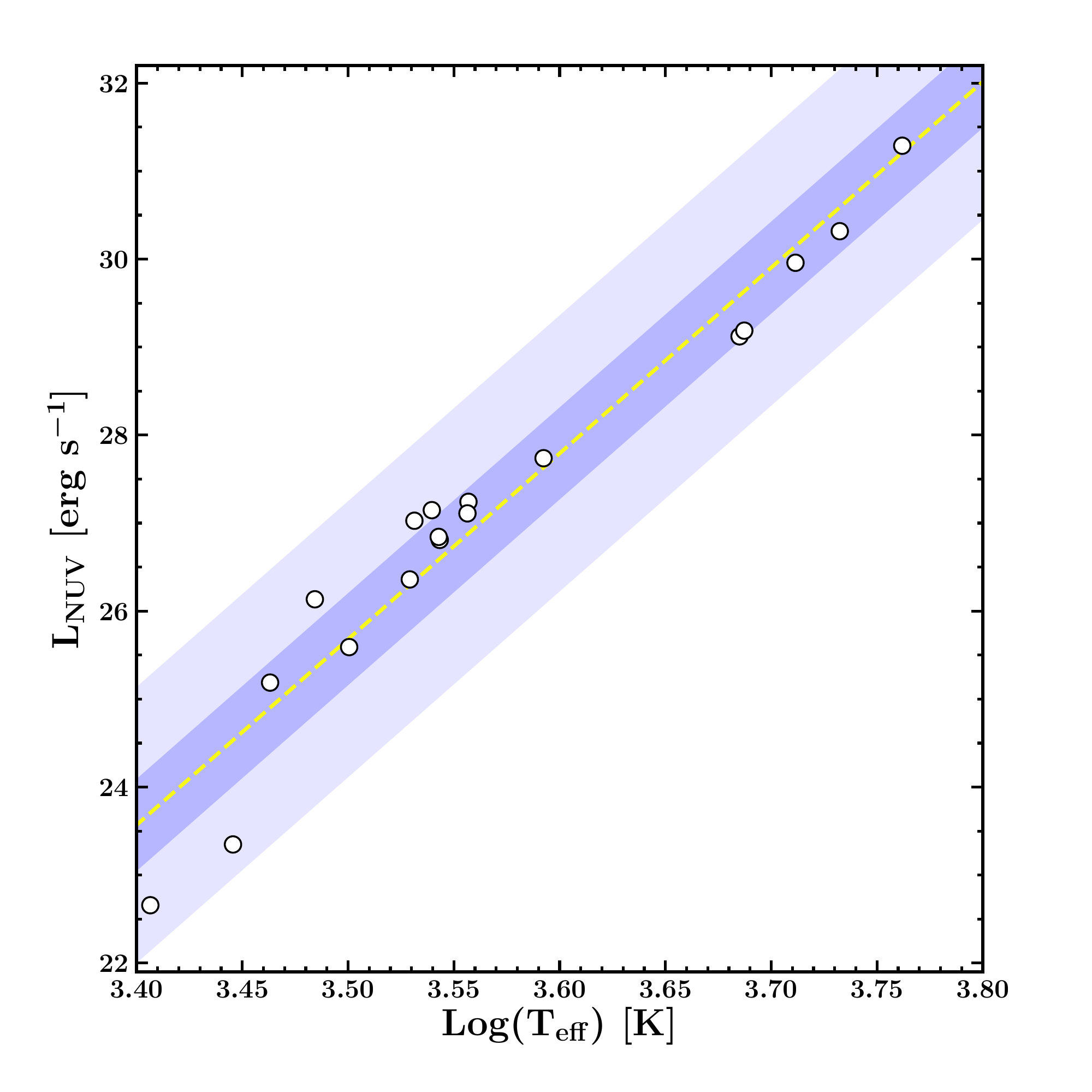}
    \caption{Correlation between  L$_{\rm NUV}$ and T$_{\rm eff}$. Open symbols represent our target sample for which we estimated $L_{\rm NUV}$ from the available ${\it Swift}$/UVOT observations. The yellow dashed line is a least-squares fit of the data points. The shaded    regions show the 1$\sigma$ and 3$\sigma$ dispersion of the data point around the best fit line. The dispersion is measured as the $\sigma$ parameter of the Gaussian distribution of the distances (computed perpendicular) of the data points from the best fit line. Proxima Centauri is the fourth source from the left.}
    \label{fig:corr}
\end{figure}

It is interesting to relate $L_{\rm NUV}$ (erg s$^{-1})$ to the stellar  $T_{\rm eff}$ (K), and estimate the CHZ from there. To this aim, we fit a log-log relation between our measured $L_{\rm NUV}$ and $T_{\rm eff}$, where the latter was obtained from the TESS Input Catalog \citep[TIC,][]{Stassun2019} for most of our targets. For Proxima Centauri, GJ 273, GJ 832, Trappist-1 and GJ 229A no $T_{\rm eff}$ was reported in the TIC and the values are respectively from \citet{Aglanda2016, Astudillo2017, Turnbull2015, Gillon2017, Gaia2018}. The $T_{\rm eff}-L_{\rm NUV}$ data are shown in Fig. \ref{fig:corr}. A simple least-squares fit produces the following log-log relation:
\begin{equation}
    \log_{10}{L_{\rm NUV}} = (21.12)\log_{10}{T_{\rm eff}}-(48.22).
    \label{eq:corr}
\end{equation}

The relation between $T_{\rm eff}$ and NUV luminosity (Eq. \ref{eq:corr} above) is a steep powerlaw (L$_{\rm NUV}$ $\propto$ T$^{21}_{\rm eff}$). This can be justified considering that for a blackbody spectrum, the ratio $F_{\rm NUV}/F_{\rm bol}$ is approximately $\propto T_{\rm eff}^{12}$ (i.e. the colder is the star the more the NUV band samples the blackbody Wien tail). Moreover,  in the range $2800\lesssim T_{\rm eff} \lesssim 6000$ K, the stellar radius $R_{\star}$ is approximately $\propto T^{2.5}_{\rm eff}$. As a result, from the Stefan-Boltzmann law ($F_{\rm bol} \propto  T_{\rm eff}^{4}$) we obtain $L_{\rm NUV} \propto F_{\rm NUV} R_{\star}^2 \propto T_{\rm eff}^{21}$. In Fig.\ref{fig:NUV} stars are colour-coded according to their $T_{\rm eff}$ (red to yellow colours going from M-type to G-type stars). 

Adopting the relations from the Baraffe stellar evolutionary models \citep{Baraffe1998}, we obtained a stellar radius-$T_{\rm eff}$ relation through which we  estimated the bolometric luminosity $L_{\rm bol}$ from the Stefan-Boltzmann law. With $T_{\rm eff}$ an $L_{\rm bol}$ at hand, we could  rely on the very definition of CHZ given by \citet{Kopparapu2014}, who indeed defined the CHZ in terms of the bolometric flux received as a function of the stellar $T_{\rm eff}$. 
With Eq.~\ref{eq:corr}, we estimated the orbital boundaries of the CHZ as a function of $T_{\rm eff}$: this is shown by the curved shaded region in   Fig.~\ref{fig:NUV}.  
The dark (light) green regions in Fig.~\ref{fig:corr} were obtained by shifting the relation given by Eq.~\ref{eq:corr} by 1-$\sigma$ (3-$\sigma$). 

Fig.~\ref{fig:NUV} shows that the agreement between the CHZ measured from the NUV luminosity (horizontal bars on each symbol) and that based on bolometric luminosity and $T_{\rm eff}$ (curved green stripes) is generally good, though some notably exceptions stand out. For instance, Proxima has a reported $T_{\rm eff}$ such that it lies above the fitted $L_{\rm NUV}$--$T_{\rm eff}$ relation of Fig.~\ref{fig:corr}, hence the two different estimates of Proxima's CHZ appear somewhat off in Fig.~\ref{fig:NUV}.

\section{Discussion}
\label{sec:discussion}

In Fig.~\ref{fig:NUV} the UV luminosity $L_{\rm NUV}$ of the host star is plotted against the planet--to--star distance $a$. Rocky planets located in the CHZ of their host stars with available UV observations are shown as open circles. For each planet, the CHZ is represented by the horizontal bar which delimits the specific range of $a$ where H$_2$O can be present in liquid phase on planet's surfaces. The shaded (blue, red, violet) diagonal stripe in Fig.~\ref{fig:NUV} represents the combination of $L_{\rm NUV}$ and $a$ where abiogenesis can actually occur and the UV irradiation is sustainable by planets' biota. 

The main conclusion that can be drawn from Fig.~\ref{fig:NUV} is that the averaged NUV luminosity of M--dwarfs is too low to trigger abiogenesis, through the considered pathways, on planets located wihin their respective CHZ (see also \citealt{2017AsBio..17..169R, Rimmer2018}). Even in the extreme case of atmospheres completely transparent to NUV radiation ($f=1.0$, grey diagonal stripe), CHZ bars do not intersect the UHZ stripe, except in the case of Proxima Centauri. Indeed, the very inner edge of the habitable zone of Proxima Centauri, which has a particularly high NUV emission given its $T_{\rm eff}$, overlaps with the UHZ zone ($f=1.0$, grey diagonal stripe), though the actual position of the Proxima Centauri b lies out of this region. We must note that planets around M--dwarfs may still be subject to abiogenesis, provided their host stars are characterized by intense and frequent NUV flare activity  \citep{gunther2019, Murray2022} or during the early stages of stellar evolution \citep{Chavali2022}.

Our analysis of UVOT observations allowed us to derive an empirical relation between the $L_{\rm NUV}$ and $T_{\rm eff}$ (Eq.\ref{eq:corr} and Fig.~\ref{fig:corr}). 
Furthermore, through stellar models (Baraffe et al. 1998) providing radii and bolometric
luminosities, we can infer the CHZ as defined in  \citet{Kopparapu2014} (shaded green curved
region in Fig.~\ref{fig:NUV}). 
The intersection between the UHZ and the CHZ is limited by a threshold temperature which excludes most M--dwarfs. According to the transparency of the atmosphere (here parametrized with $f$ which represents the fraction of absorbed flux), we find that even for the completely transparent case ($f=1.0$) stars with  T$_{eff}\gtrsim 3900$ K may have experienced the correct NUV flux (sufficient for abiogensis and not too high to be dangerous for biota). Such thredshold temperature increases to $T_{\rm eff}\gtrsim 4200$ K if $f=0.5$ (red shaded region) and  T$_{eff}\gtrsim4700$ K if  $f=0.1$ (violet shaded region). 

It is worth noting here that we neglected any transformation of the atmosphere induced by the absorbed UV radiation. As an example, let's consider an atmosphere fully opaque to UV radiation i.e., the planetary surface is shielded from energetic photons. This can be positive for present-life, but negative for the possibility of UV-triggered abiogenesis. At the same time the atmosphere could be transformed over time by the absorbed UV radiation, possibly driving an evolution of the transmission properties over time, and then an evolution of the UHZ for a specific planet. A detailed discussion of these issues is beyond the scope of this work.


We must also note that we used the $L_{\rm NUV}-T_{\rm eff}$ relation reported in Eq.~\ref{eq:corr}  only for the purpose of representing the H$_2$O habitable zone in Fig.~\ref{fig:NUV}. Indeed, the steepness of the correlation prevents its use to estimate the NUV luminosity given T$_{\rm eff}$, a small difference in temperature can lead to orders of magnitudes differences in $L_{\rm NUV}$. This owns a particular relevance in view of the different estimates of $T_{\rm eff}$ often present in the literature (up to hundreds degrees) for the very same star (see, e.g., the NASA Exoplanet Archive\footnote{\url{https://exoplanetarchive.ipac.caltech.edu/}}). 
It is therefore clear that direct NUV data are necessary.

In our derivation of the UHZ shown in Fig.~\ref{fig:NUV} we considered the flux threshold needed to trigger abiogenesis proposed by \citet{Rimmer2018}. This is the flux leading to a 50\% yield of the photochemical products at 0$^{\circ}$C in the SO$^{2-}_{3}$ reaction (\S \ref{sec:habitablezone}). As discussed by \citet{Rimmer2018} this choice is very liberal, since any prebiotic synthesis will take several steps, and even a 50\% yield per step will probably result in too small a final concentration of life's building blocks to do anything (see the discussion of the "arithmetic demon" in \citealt{Rimmer2018}). 
Relaxing the \% yield will only work if prebiotic chemists find a way to stepwise concentrate preferentially low yield products.

In this case, if we allow for lower yields, i.e., for a lower flux value, the abiogenesis limit will  consequently shift to the right in Fig.~\ref{fig:NUV}. 
As an example, at 0$^{\circ}$C, the flux required to obtain a 25\% and a 10\% yield is a factor of 4 and 20 less than that required to obtain a 50\% yield (see Fig.~3 in \citealt{Rimmer2018}). On the contrary, if we assume a larger temperature, the abiogenesis limit would shift to the left, since a higher flux is required for the same yield at higher temperatures (see Fig. 3 in \citealt{Rimmer2018}). As an example, at 20$^{\circ}$C and at 40$^{\circ}$C the NUV flux required is $\sim$25 and $\sim$250 times higher than that required at 0$^{\circ}$C to have a 50\% yield. The right boundary of the UHZ would scale linearly with these factors.

In order to estimate the inner boundary of UHZ (\S \ref{sec:habitablezone}), inspired by the Principle of Mediocrity we assumed the maximum tolerable NUV flux as twice the flux experienced by the Archean Earth above the atmosphere.
If life can resist a larger NUV flux, then a planet can orbit closer to its star. This corresponds to shifting the limits of the UHZ zone in Fig. \ref{fig:NUV} to the left.

In our work we did not consider possible effects caused by the Far-UV radiation (FUV, <200 nm). While FUV radiation is important in the context of atmospheric photochemistry, it is most probably unable to reach the surface of habitable planets because of severe absorption by atmospheric constituents, such as water vapor and carbon dioxide \citep{Cnossen2007, 2017ApJ...843..110R}. 

In is important to realize that the prebiotic pathways we considered are not the only possible for explaining the origin of life \citep[e.g.,][]{Cleaves2008, Ruiz2013}. 
What is particularly intriguing is that the UV-criterion for abiogenesis was derived through  laboratory experiments, that showed how much UV radiation is required to form RNA pyrimidines
\citep{Powner2009, Ritson2012, Patel2015, Xu2018, Rimmer2018}. This finding is supported also by the fact that RNA exhibits selection pressure under UV radiation, suggesting that it is formed in a UV-rich environment \citep{Rios2013, Beckstead2016, Pollum2016}. Moreover, as shown by \citet{Deamer2010}, UV radiation was by far the most abundant source of chemical free energy present on the surface of the early Earth.

As suggested by \citet{Lillo2022} K-type stars may currently be the best candidates for searching planets with suitable conditions for life as we know it. Due to their closer CHZ, the CHZ planet detection via transit amd/or radial velocity is easier compared to G-stars. Moreover, K-type stars are longer-lived with respect to G-star and thus they offer
longer timescales for biological evolution on their orbiting planets. In contrast with M-dwarfs, the lower activity/variability of K-stars favours detectability and habitability. Furthermore, compared to M--dwarfs, the habitable planets around K-stars are not tidal-locked \citep{Heller2011, Barnes2017}. In our study, we further demonstrated that K-stars can irradiate planets in the CHZ with a NUV flux sufficient to trigger abiogenesis.

In the next decades exoplanetary research will be able to carry out statistical studies on the presence of biosignatures in the atmosphere of exoplanets. This brings about a change of perspective, turning exoplanets into laboratories for testing theories about the origin of life. Should we find life on a planet around M--dwarfs, we will have to imagine another mechanisms able to provide the necessary UV flux in these stars (e.g, flares or early stages of stellar evolution), or we will have to reject the prebiotic pathways to life we considered here.

\section*{Acnowledgements}
This study made use of data supplied by the UK {\it Swift} Science Data Centre at the University of Leicester. Part of this work is based on archival data, software, or online services provided by the Space Science Data Center-ASI. RS acknowledges Sergio Campana and Claudia Scarlata for the suggestions regarding the Swift-UVOT data analysis and Jacopo Pravettoni for the valuable support for data visualization. We also thank the referee, Paul Rimmer, for the helpful comments and suggestions that improved the quality of the paper.

\section*{Data Availability}
The data underlying this article are available in the Swift Data Archive Mirror, see: {\url{https://www.ssdc.asi.it/mmia/index.php?mission=swiftmastr}}.



\bibliographystyle{mnras}

\begin{thebibliography}{}

\bibitem[Anglada-Escud{\'e} et al.(2016)]{Aglanda2016} Anglada-Escud{\'e}, G., Amado, P.~J., Barnes, J., et al.\ 2016, \nat, 536, 437. doi:10.1038/nature19106
\bibitem[Astudillo-Defru et al.(2017)]{Astudillo2017} Astudillo-Defru, N., Forveille, T., Bonfils, X., et al.\ 2017, \aap, 602, A88. doi:10.1051/0004-6361/201630153
\bibitem[Baraffe et al.(1998)]{Baraffe1998} Baraffe, I., Chabrier, G., Allard, F., et al.\ 1998, \aap, 337, 403
\bibitem[Barnes(2017)] {Barsen2017} Barnes, R.\ 2017, Celestial Mechanics and Dynamical Astronomy, 129, 509 
\bibitem[Barclay et al.(2021)]{Barclay2021} Barclay, T., Kostov, V.~B., Col{\'o}n, K.~D., et al.\ 2021, \aj, 162, 300. doi:10.3847/1538-3881/ac2824
\bibitem[Barstow \& Irwin(2016)]{Barstow2016} Barstow, J.~K. \& Irwin, P.~G.~J.\ 2016, \mnras, 461, L92. doi:10.1093/mnrasl/slw109
\bibitem[Barnes(2017)]{Barnes2017} Barnes, R.\ 2017, Celestial Mechanics and Dynamical Astronomy, 129, 509. doi:10.1007/s10569-017-9783-7
\bibitem[Batalha et al.(2015)]{Batalha2015} Batalha, N., Kalirai, J., Lunine, J., et al.\ 2015, arXiv:1507.02655
\bibitem[Beckstead et al.(2016)]{Beckstead2016} Beckstead, A. A., Zhang, Y., de Vries, M. S., \& Kohler, B. 2016, Physical Chemistry Chemical Physics, 18, 24228
\bibitem[Benedict et al.(2016)]{Benedict2016} Benedict, G.~F., Henry, T.~J., Franz, O.~G., et al.\ 2016, \aj, 152, 141 
\bibitem[Benneke et al.(2019)]{Benneke2019} Benneke, B., Wong, I., Piaulet, C., et al.\ 2019, \apjl, 887, L14. doi:10.3847/2041-8213/ab59dc
\bibitem[Bibring et al.(2006)]{Bibring2006} Bibring, J.-P., Langevin, Y., Mustard, J.~F., et al.\ 2006, Science, 312, 400. doi:10.1126/science.1122659

\bibitem[Breeveld et al.(2011)]{Breeveld2011} Breeveld, A.~A., Landsman, W., Holland, S.~T., et al.\ 2011, Gamma Ray Bursts 2010, 1358, 373. doi:10.1063/1.3621807
\bibitem[Brown et al.(2010)]{Brown2010} Brown, P.~J., Roming, P.~W.~A., Milne, P., et al.\ 2010, \apj, 721, 1608. doi:10.1088/0004-637X/721/2/1608
\bibitem[Brown et al.(2016)]{Brown2016} Brown, P.~J., Breeveld, A., Roming, P.~W.~A., et al.\ 2016, VizieR Online Data Catalog, J/AJ/152/102
\bibitem[Buccino et al.(2007)]{Buccino2007} Buccino, A.~P., Lemarchand, G.~A., \& Mauas, P.~J.~D.\ 2007, \icarus, 192, 582 
\bibitem[Burrows et al.(2005)]{Burrows2005} Burrows, D.~N., Hill, J.~E., Nousek, J.~A., et al.\ 2005, \ssr, 120, 165. doi:10.1007/s11214-005-5097-2
\bibitem[Bochanski et al.(2010)]{Bochansky2010} Bochanski, J.~J., Hawley, S.~L., Covey, K.~R., et al.\ 2010, \aj, 139, 2679 

\bibitem[Catling et al.(2018)]{Catling2018} Catling, D.~C., Krissansen-Totton, J., Kiang, N.~Y., et al.\ 2018, Astrobiology, 18, 709. doi:10.1089/ast.2017.1737
\bibitem[Chavali et al.(2022)]{Chavali2022} Chavali, S., Youngblood, A., Paudel, R.~R., et al.\ 2022, Research Notes of the American Astronomical Society, 6, 201. doi:10.3847/2515-5172/ac9591
\bibitem[Claire et al.(2012)]{Claire2012} Claire, M.~W., Sheets, J., Cohen, M., et al.\ 2012, \apj, 757, 95 
\bibitem[Cloutier et al.(2017)]{Cloutier2017} Cloutier, R., Astudillo-Defru, N., Doyon, R., et al.\ 2017, \aap, 608, A35. doi:10.1051/0004-6361/201731558
\bibitem[Cleaves et al.(2008)]{Cleaves2008}
Cleaves HJ, Chalmers JH, Lazcano A, Miller SL, Bada JL,  2008, Orig Life Evol Biosph, 105-15. 
\bibitem[Cnossen et al.(2007)]{Cnossen2007} Cnossen, I., Sanz-Forcada, J., Favata, F., et al.\ 2007, Journal of Geophysical Research (Planets), 112, E02008. doi:10.1029/2006JE002784
\bibitem[Cockell(1998)]{Cockell1998} Cockell, C.~S.\ 1998, Journal of Theoretical Biology, 193, 717. doi:10.1006/jtbi.1998.0738
\bibitem[Cockell(2000)]{Cockell2000} Cockell, C.~S.\ 2000, \planss, 48, 203. doi:10.1016/S0032-0633(99)00087-2

\bibitem[Deamer \& Weber(2010)]{Deamer2010} Deamer D and Weber AL, 2010, Bioenergetics and life's origins. Cold Spring Harb Perspect Biol 2:a004929.

\bibitem[Del Genio et al.(2018)]{DelGenio2018} Del Genio, A.~D., Kiang, N.~Y., Way, M.~J., et al.\ 2018, arXiv:1812.06606 
\bibitem[Des Marais et al.(2002)]{Desmarais2002} Des Marais, D.~J., Harwit, M.~O., Jucks, K.~W., et al.\ 2002, Astrobiology, 2, 153. doi:10.1089/15311070260192246
\bibitem[de Zeeuw et al.(2014)]{dezeeuw2014} de Zeeuw, T., Tamai, R., \& Liske, J.\ 2014, The Messenger, 158, 3

\bibitem[Dittmann et al.(2017)]{Dittmann} Dittmann, J.~A., Irwin, J.~M., Charbonneau, D., et al.\ 2017, \nat, 544, 333 
\bibitem[Dodd et al.(2017)]{Dodd2017} Dodd, M.~S., Papineau, D., Grenne, T., et al.\ 2017, \nat, 543, 60. doi:10.1038/nature21377
\bibitem[Dreizler et al.(2020)]{Dreizler2020} Dreizler, S., Jeffers, S.~V., Rodr{\'\i}guez, E., et al.\ 2020, \mnras, 493, 536. doi:10.1093/mnras/staa248

\bibitem[Dressing \& Charbonneau(2015)]{2015ApJ...807...45D} Dressing, C.~D., \& Charbonneau, D.\ 2015, \apj, 807, 45 
\bibitem[Driscoll \& Barnes(2015)]{Driscoll2015} Driscoll, P.~E. \& Barnes, R.\ 2015, Astrobiology, 15, 739. doi:10.1089/ast.2015.1325

\bibitem[Feng et al.(2018)]{2018arXiv180702483F} Feng, F., Tuomi, M., \& Jones, H.~R.~A.\ 2018, arXiv:1807.02483
\bibitem[Foley(2019)]{Foley2019} Foley, B.~J.\ 2019, \apj, 875, 72. doi:10.3847/1538-4357/ab0f31
\bibitem[Ferris \& Orgel(1966)]{Ferris1966} Ferris J.P.,  Orgel L. E., 1966 J. Am. Chem. Soc. , 88, 5, 1074
\bibitem[Foreman-Mackey et al.(2015)]{Foreman2015} Foreman-Mackey, D., Montet, B.~T., Hogg, D.~W., et al.\ 2015, \apj, 806, 215. doi:10.1088/0004-637X/806/2/215
\bibitem[France et al.(2012)]{France2012} France, K., Linsky, J.~L., Tian, F., Froning, C.~S., \& Roberge, A.\ 2012, \apjl, 750, L32 
\bibitem[France et al.(2013)]{2013ApJ...763..149F} France, K., Froning, C.~S., Linsky, J.~L., et al.\ 2013, \apj, 763, 149
\bibitem[France et al.(2016)]{2016ApJ...820...89F} France, K., Loyd, R.~O.~P., Youngblood, A., et al.\ 2016, \apj, 820, 89.
\bibitem[Gaia Collaboration et al.(2018)]{Gaia2018} Gaia Collaboration, Brown, A.~G.~A., Vallenari, A., et al.\ 2018, \aap, 616, A1
\bibitem[Gardner et al.(2006)]{Gardner2006} Gardner, J.~P., Mather, J.~C., Clampin, M., et al.\ 2006, \ssr, 123, 485. doi:10.1007/s11214-006-8315-7
\bibitem[Gehrels et al.(2004)]{Gehrels2004} Gehrels, N., Chincarini, G., Giommi, P., et al.\ 2004, \apj, 611, 1005 
\bibitem[Gilbert(1986)]{Gilbert1986} Gilbert, W.\ 1986, \nat, 319, 618. doi:10.1038/319618a0
\bibitem[Gillon et al.(2017)]{Gillon2017} Gillon, M., Triaud, A.~H.~M.~J., Demory, B.-O., et al.\ 2017, \nat, 542, 456. doi:10.1038/nature21360
\bibitem[Gilmozzi \& Spyromilio(2007)]{Gilmozzi2007} Gilmozzi, R. \& Spyromilio, J.\ 2007, The Messenger, 127, 11
\bibitem[Godolt et al.(2019)]{Godolt2019} Godolt, M., Tosi, N., Stracke, B., et al.\ 2019, \aap, 625, A12. doi:10.1051/0004-6361/201834658
\bibitem[Guinan et al.(2016)]{2016ApJ...821...81G} Guinan, E.~F., Engle, S.~G., \& Durbin, A.\ 2016, \apj, 821, 81 
\bibitem[G{\"u}nther et al.(2019)]{gunther2019} G{\"u}nther, M.~N., Pozuelos, F.~J., Dittmann, J.~A., et al.\ 2019, Nature Astronomy, 3, 1099. doi:10.1038/s41550-019-0845-5
\bibitem[Harman et al.(2015)]{2015ApJ...812..137H} Harman, C.~E., Schwieterman, E.~W., Schottelkotte, J.~C., \& Kasting, J.~F.\ 2015, \apj, 812, 137 
\bibitem[Heath et al.(1999)]{1999OLEB...29..405H} Heath, M.~J., Doyle, L.~R., Joshi, M.~M., \& Haberle, R.~M.\ 1999, Origins of Life and Evolution of the Biosphere, 29, 405
\bibitem[Heller et al.(2011)]{Heller2011} Heller, R., Leconte, J., \& Barnes, R.\ 2011, \aap, 528, A27. doi:10.1051/0004-6361/201015809

\bibitem[Kane(2018)]{2018ApJ...861L..21K} Kane, S.~R.\ 2018, \apjl, 861, L21
\bibitem[Kasting(1988)]{Kasting1988} Kasting, J.~F.\ 1988, \icarus, 74, 472. doi:10.1016/0019-1035(88)90116-9
\bibitem[Kasting(1991)]{Kasting1991} Kasting, J.~F.\ 1991, \icarus, 94, 1. doi:10.1016/0019-1035(91)90137-I
\bibitem[Kasting et al.(1993)]{Kasting1993} Kasting, J.~F., Whitmire, D.~P., \& Reynolds, R.~T.\ 1993, \icarus, 101, 108 

\bibitem[Kopparapu et al.(2013)]{Kopparapu2013} Kopparapu, R.~K., Ramirez, R., Kasting, J.~F., et al.\ 2013, \apj, 765, 131 

\bibitem[Kopparapu et al.(2014)]{Kopparapu2014} Kopparapu, R.~K., Ramirez, R.~M., SchottelKotte, J., et al.\ 2014, \apjl, 787, L29. doi:10.1088/2041-8205/787/2/L29

\bibitem[Lakatos (1974)]{Lakatos1974} Lakatos, I., 1974, in Criticism and the Growth of Knowledge, “Falsification and the methodology of scientific research programs”, eds Lakatos, I., \& Musgrave, A., Cambridge University, Press, Cambridge, p. 91


\bibitem[Lammer et al.(2009)]{2009A&ARv..17..181L} Lammer, H., Bredeh{\"o}ft, J.~H., Coustenis, A., et al.\ 2009, \aapr, 17, 181 
\bibitem[Lenardic et al.(2014)]{Lenardic2014} Lenardic, A., Hoeink, T., Jellinek, M., et al.\ 2014, AGU Fall Meeting Abstracts
\bibitem[Lillo-Box et al.(2022)]{Lillo2022} Lillo-Box, J., Santos, N.~C., Santerne, A., et al.\ 2022, arXiv:2209.05205

\bibitem[Lingam \& Loeb(2017)]{2017PNAS..114.6689L} Lingam, M., \& Loeb, A.\ 2017, Proceedings of the National Academy of Science, 114, 6689 
\bibitem[Linsky(2014)]{2014Chall...5..351L} Linsky, J.\ 2014, Challenges, 5, 351 
\bibitem[Loyd et al.(2018)]{2018ApJ...867...71L} Loyd, R.~O.~P., France, K., Youngblood, A., et al.\ 2018, \apj, 867, 71 


\bibitem[Luger \& Barnes(2015)]{2015AsBio..15..119L} Luger, R., \& Barnes, R.\ 2015, Astrobiology, 15, 119
\bibitem[Luque et al.(2019)]{Luque2019} Luque, R., Pall{\'e}, E., Kossakowski, D., et al.\ 2019, \aap, 628, A39. doi:10.1051/0004-6361/201935801

\bibitem[Lustig-Yaeger et al.(2019)]{Lusting2019} Lustig-Yaeger, J., Lincowski, A., \& Meadows, V.\ 2019, The American Astronomical Society

\bibitem[Meadows et al.(2017)]{Meadows2017} Meadows, V., Arney, G., Schwieterman, E., et al.\ 2017, American Astronomical Society Meeting Abstracts, 229, 120.03 

\bibitem[Meadows \& Barnes(2018)]{Meadows2018} Meadows, V.~S. \& Barnes, R.~K.\ 2018, Handbook of Exoplanets, 57. doi:10.1007/978-3-319-55333-7\_57
\bibitem[McIntyre(2022)]{McIntyre2022} McIntyre, S.~R.~N.\ 2022, \aap, 662, A15. doi:10.1051/0004-6361/202141112

\bibitem[Melott \& Thomas(2011)]{Melott2011} Melott, A.~L. \& Thomas, B.~C.\ 2011, Astrobiology, 11, 343
\bibitem[Ment et al.(2018)]{ment} Ment, K., Dittmann, J.~A., Astudillo-Defru, N., et al.\ 2018, arXiv:1808.00485 
\bibitem[Miller (1957)]{Miller1957} Biochimica et Biophysica Acta,23, 480-489
\bibitem[Miles \& Shkolnik(2017)]{Miles2017} Miles, B.~E., \& Shkolnik, E.~L.\ 2017, \aj, 154, 67 
\bibitem[Montet et al.(2015)]{Montet2015} Montet, B.~T., Morton, T.~D., Foreman-Mackey, D., et al.\ 2015, \apj, 809, 25.
doi:10.1088/0004-637X/809/1/25
\bibitem[Morrissey et al.(2007)]{Morrissey2007} Morrissey, P., Conrow, T., Barlow, T.~A., et al.\ 2007, \apjs, 173, 682.
\bibitem[Murray et al.(2022)]{Murray2022} Murray, C.~A., Queloz, D., Gillon, M., et al.\ 2022, \mnras, 513, 2615. doi:10.1093/mnras/stac1078
\bibitem[Newton et al.(2018)]{Newton2018} Newton, E.~R., Mondrik, N., Irwin, J., Winters, J.~G., \& Charbonneau, D.\ 2018, arXiv:1807.09365 
\bibitem[Nutzman \& Charbonneau(2008)]{Nutzman2008} Nutzman, P., \& Charbonneau, D.\ 2008, \pasp, 120, 317 
\bibitem[Owen(1980)]{Owen1980} Owen, T.\ 1980, Strategies for the Search for Life in the Universe, 83, 177. doi:10.1007/978-94-009-9115-6\_17
\bibitem[Patel et al.(2015)]{Patel2015} Patel, B.~H., Percivalle, C., Ritson, D.~J., et al.\ 2015, Nature Chemistry, 7, 301. doi:10.1038/nchem.2202

\bibitem[Paturel et al.(2003)]{2003A&A...412...45P} Paturel, G., Petit, C., Prugniel, P., et al.\ 2003, \aap, 412, 45 
\bibitem[Pickles(1998)]{Pickles1998} Pickles, A.~J.\ 1998, \pasp, 110, 863. doi:10.1086/316197
\bibitem[Piran \& Jimenez(2014)]{Piran2014} Piran, T. \& Jimenez, R.\ 2014, \prl, 113, 231102
\bibitem[Powner et al.(2009)]{Powner2009} Powner, M.~W., Gerland, B., \& Sutherland, J.~D.\ 2009, \nat, 459, 239. doi:10.1038/nature08013
\bibitem[Pollum et al. (2016)]{Pollum2016} Pollum, M., Ashwood, B., Jockusch, S., Lam, M., \&
Crespo-Hern´andez, C. E. 2016, Journal of the American Chemical Society, 138, 11457
\bibitem[Pollack et al.(1987)]{Pollack1987} Pollack, J.~B., Kasting, J.~F., Richardson, S.~M., et al.\ 1987, \icarus, 71, 203. doi:10.1016/0019-1035(87)90147-3

\bibitem[Quirrenbach et al.(2014)]{Quirrenbach2014} Quirrenbach, A., Amado, P.~J., Caballero, J.~A., et al.\ 2014, Exploring the Formation and Evolution of Planetary Systems, 299, 395 
\bibitem[Ranjan et al.(2017)]{2017ApJ...843..110R} Ranjan, S., Wordsworth, R., \& Sasselov, D.~D.\ 2017, \apj, 843, 110
\bibitem[Ranjan \& Sasselov(2017)]{2017AsBio..17..169R} Ranjan, S. \& Sasselov, D.~D.\ 2017, Astrobiology, 17, 169. doi:10.1089/ast.2016.1519
\bibitem[Ramirez \& Kaltenegger(2017)]{Ramirez2017} Ramirez, R.~M. \& Kaltenegger, L.\ 2017, \apjl, 837, L4. doi:10.3847/2041-8213/aa60c8
\bibitem[Redfield \& Linsky(2000)]{2000ApJ...534..825R} Redfield, S., \& Linsky, J.~L.\ 2000, \apj, 534, 825 
\bibitem[Redfield \& Linsky(2008)]{2008ApJ...673..283R} Redfield, S., \& Linsky, J.~L.\ 2008, \apj, 673, 283
\bibitem[Ribas et al.(2016)]{2016A&A...596A.111R} Ribas, I., Bolmont, E., Selsis, F., et al.\ 2016, \aap, 596, A111 
\bibitem[Rich \& Kasha (1962)]{Rich1962} Rich, A., \& Kasha, M. (1962). Horizons in biochemistry. Eds. M. Kasha and B. Pullman, Academic Press, New York, 103
\bibitem [Rimmer et al.(2018)] {Rimmer2018} Rimmer, Paul B. and Xu, Jianfeng and Thompson, Samantha J. and Gillen, Ed and Sutherland, John D. and Queloz, Didier, The origin of RNA precursors on exoplanets, 2018, American Association for the Advancement of Science
\bibitem[\protect\citeauthoryear{Rimmer et al.}{2021}]{Rimmer2021} Rimmer P.~B., Thompson S.~J., Xu J., Russell D.~A., Green N.~J., Ritson D.~J., Sutherland J.~D., et al., 2021, AsBio, 21, 1099. doi:10.1089/ast.2020.2335

\bibitem[Rios \& Tor(2013)]{Rios2013} Rios, A. C., \& Tor, Y. 2013, Israel journal of chemistry, 53,
469
\bibitem[Ritson \& Sutherland(2012)]{Ritson2012} Ritson, D. \& Sutherland, J.~D.\ 2012, Nature Chemistry, 4, 895. doi:10.1038/nchem.1467
\bibitem[Rodler \& L{\'o}pez-Morales(2014)]{Rodler2014} Rodler, F. \& L{\'o}pez-Morales, M.\ 2014, \apj, 781, 54. doi:10.1088/0004-637X/781/1/54
\bibitem[Roming et al.(2005)]{Roming2005} Roming, P.~W.~A., Kennedy, T.~E., Mason, K.~O., et al.\ 2005, \ssr, 120, 95. doi:10.1007/s11214-005-5095-4

\bibitem[Ruiz-Mirazo et al(2013)]{Ruiz2013} Rui-Mirazo K., Briones C., de la Escosura A., 2013, Chem. Rev. 2014, 114, 1, 285–366

\bibitem[Ruderman(1974)]{Ruderman1974} Ruderman, M.~A.\ 1974, Science, 184, 1079
\bibitem [Sagan(1973)]{Sag} Sagan, C. 1973, Journal of theoretical biology, 39, 195
\bibitem[Sagan et al.(1993)]{Sagan1993} Sagan, C., Thompson, W.~R., Carlson, R., Gurnett, D., Hord, C.\ 1993.\ A search for life on Earth from the Galileo spacecraft.\ Nature 365, 715–721
\bibitem[Sanchez et al.(1966)]{Sanchez1966} Sanchez, R.~A., Ferris, J.~P., \& Orgel, L.~E.\ 1966, Science, 154, 784. doi:10.1126/science.154.3750.784
\bibitem[Sanchez \& Orgel(1970)]{Sanchez1970} Sanchez, RA. Orgel, LE. Studies in prebiotic synthesis. V. Synthesis and photoanomerization of pyrimidine nucleosides. J Mol Biol. 47, 531-43, 1970
\bibitem[Sanz-Forcada et al.(2010)]{SanzForcada2010} Sanz-Forcada, J., Ribas, I., Micela, G., et al.\ 2010, \aap, 511, L8 

\bibitem[Schneider \& Shkolnik(2018)]{2018AJ....155..122S} Schneider, A.~C., \& Shkolnik, E.~L.\ 2018, \aj, 155, 122 
\bibitem[Schwieterman et al.(2018)]{Schwieterman2018} Schwieterman, E.~W., Kiang, N.~Y., Parenteau, M.~N., et al.\ 2018, Astrobiology, 18, 663 
\bibitem[Selsis et al.(2007)]{2007A&A...476.1373S} Selsis, F., Kasting, J.~F., Levrard, B., et al.\ 2007, \aap, 476, 1373 
\bibitem [Setlow (1974)]{Set} Setlow, R. B. 1974, Proceedings of the National Academy
of Sciences, 71, 3363
\bibitem[Shields et al.(2016)]{Shields2016} Shields, A.~L., Ballard, S., \& Johnson, J.~A.\ 2016, \physrep, 663, 1 
\bibitem[Snellen et al.(2013)]{Snellen2013} Snellen, I.~A.~G., de Kok, R.~J., le Poole, R., et al.\ 2013, \apj, 764, 182. doi:10.1088/0004-637X/764/2/182
\bibitem[Solomon \& Head(1991)]{Solomon1991} Solomon, S.~C. \& Head, J.~W.\ 1991, Science, 252, 252. doi:10.1126/science.252.5003.252
\bibitem[Spinelli et al.(2021)]{Spinelli2021} Spinelli, R., Ghirlanda, G., Haardt, F., et al.\ 2021, \aap, 647, A41. doi:10.1051/0004-6361/202039507
\bibitem[Stassun et al.(2019)]{Stassun2019} Stassun, K.~G., Oelkers, R.~J., Paegert, M., et al.\ 2019, \aj, 158, 138. doi:10.3847/1538-3881/ab3467
\bibitem[Stelzer et al.(2013)]{Stelzer2013} Stelzer, B., Marino, A., Micela, G., L{\'o}pez-Santiago, J., \& Liefke, C.\ 2013, \mnras, 431, 2063 
\bibitem[Stelzer et al.(2016)]{2016MNRAS.463.1844S} Stelzer, B., Damasso, M., Scholz, A., \& Matt, S.~P.\ 2016, \mnras, 463, 1844 
\bibitem[Tarter et al.(2007)]{2007AsBio...7...30T} Tarter, J.~C., Backus, P.~R., Mancinelli, R.~L., et al.\ 2007, Astrobiology, 7, 30 
\bibitem[Takai et al.(2008)]{Takai2008} Takai, K., Nakamura, K., Toki, T., et al.\ 2008, Proceedings of the National Academy of Science, 105, 10949. doi:10.1073/pnas.0712334105
\bibitem[Thorsett(1995)]{Thorsett1995} Thorsett, S.~E.\ 1995, \apjl, 444, L53
\bibitem[Thuillier et al.(2004)]{Thuillier2004} Thuillier, G., Floyd, L., Woods, T.~N., et al.\ 2004, Advances in Space Research, 34, 256. doi:10.1016/j.asr.2002.12.004

\bibitem[Tian et al.(2014)]{2014E&PSL.385...22T} Tian, F., France, K., Linsky, J.~L., Mauas, P.~J.~D., \& Vieytes, M.~C.\ 2014, Earth and Planetary Science Letters, 385, 22 
\bibitem[Tosi et al.(2017)]{Tosi2017} Tosi, N., Godolt, M., Stracke, B., et al.\ 2017, \aap, 605, A71. doi:10.1051/0004-6361/201730728

\bibitem[Toupance et al.(1977)]{Toupance1977} Toupance, G., Bossard, A., \& Raulin, F.\ 1977, Origins of Life, 8, 259. doi:10.1007/BF00930687
\bibitem[Turnbull(2015)]{Turnbull2015} Turnbull, M.~C.\ 2015, arXiv:1510.01731
\bibitem[Van Eylen et al.(2021)]{VanEylen2021} Van Eylen, V., Astudillo-Defru, N., Bonfils, X., et al.\ 2021, \mnras, 507, 2154. doi:10.1093/mnras/stab2143
\bibitem[von Hoerner(1961)]{Vonhoerner1961} von Hoerner, S.\ 1961, Science, 134, 1839. doi:10.1126/science.134.3493.1839



\bibitem[Walker et al.(1981)]{Walker1981} Walker, J.~C.~G., Hays, P.~B., \& Kasting, J.~F.\ 1981, \jgr, 86, 9776. doi:10.1029/JC086iC10p09776


\bibitem[Wheatley et al.(2017)]{2017MNRAS.465L..74W} Wheatley, P.~J., Louden, T., Bourrier, V., Ehrenreich, D., \& Gillon, M.\ 2017, \mnras, 465, L74 
\bibitem[Winters et al.(2017)]{2017yCat..51530014W} Winters, J.~G., Sevrinsky, R.~A., Jao, W.-C., et al.\ 2017, VizieR Online Data Catalog, 515, 
\bibitem[Woods et al.(2009)]{2009GeoRL..36.1101W} Woods, T.~N., Chamberlin, P.~C., Harder, J.~W., et al.\ 2009, \grl, 36, L01101 
\bibitem[Wordsworth \& Pierrehumbert(2014)]{2014ApJ...785L..20W} Wordsworth, R., \& Pierrehumbert, R.\ 2014, \apjl, 785, L20 
\bibitem[Xu et al.(2018)]{Xu2018} 
Xu, J., Ritson, D. J., Ranjan,S.,  Todd, Z. R., Sasselov, D. D., Sutherland, J. D.,\ 2018, Chem. Commun., 54, 44
\bibitem[Zahnle \& Catling(2017)]{Zahnle2017} Zahnle, K.~J. \& Catling, D.~C.\ 2017, \apj, 843, 122. doi:10.3847/1538-4357/aa7846
\bibitem[Zechmeister \& K{\"u}rster(2009)]{2009A&A...496..577Z} Zechmeister, M., \& K{\"u}rster, M.\ 2009, \aap, 496, 577 
\bibitem[Zsom et al.(2013)]{Zsom2013} Zsom, A., Seager, S., de Wit, J., et al.\ 2013, \apj, 778, 109. doi:10.1088/0004-637X/778/2/109

\end{thebibliography}




\appendix

\appendix


\bsp	
\label{lastpage}
\end{document}